\documentstyle[preprint,epsf,prb,aps]{revtex}

\begin{document}
\newlength{\figwidth}
\setlength{\figwidth}{\textwidth}
\addtolength{\figwidth}{-.25\figwidth}

\draft
\tighten
\title{A unified approach to electron transport in double barrier structures}
\author{Giuseppe Iannaccone\cite{email} and Bruno Pellegrini}
\address{Dipartimento di Ingegneria dell'Informazione: Elettronica, Informatica
e Telecomunicazioni \\
Universit\`a degli studi di Pisa, Via Diotisalvi 2, I-56126 Pisa, Italy}
\date{To be published on Phys. Rev. B - cond-mat/9510169}
\maketitle

\begin{abstract}
In this paper we show an approach to electron transport in double
barrier structures which unifies the well known sequential and
resonant tunneling models in the widest range of transport regimes,
from completely  coherent to completely incoherent.
In doing so, we make a clear distinction between ``approaches''
and ``transport regimes,'' in order to clarify some ambiguities in
the concept of sequential tunneling.
Scattering processes in the well are accounted for by means of an
effective mean free path,
which plays the role of a relaxation length.
Our approach is based on a recently derived formula for the
density of states in a quantum well, as a function of the round trip
time in the well and of trasmission and reflection probabilities for
the whole structure and for each barrier.
\end{abstract}

\pacs{PACS numbers: 71.20.-b, 73.20.Dx, 73.40.Gk}
\section{Introduction}

Tunneling in double barrier structures has
been extensively studied since the pioneering
work of Tsu, Esaki and Chang.\cite{tsuesaki73,chanesak74}
These structures promise many interesting device
applications\cite{capasso90}
and allow to study problems, relating energy levels in
quantum wells and tunneling, which are of general interest in
condensed matter physics.

A proper description of electron transport in double barriers
is required to understand the relevant phenomena affecting the electrical
properties of these structures and to construct a suitable model for
obtaining DC characteristics, high frequency performances, and
noise properties in agreement with the experimental measurements.

There are two well known models for transport through a
double barrier. The first has been basically proposed
by Chang,\cite{chanesak74} and is the ``resonant tunneling'' model,
which is easy to understand by means of the analogy with a Fabry-Perot
resonator: the tunneling probability for the whole structure is
resonantly enhanced for incident electron energies close to
the discrete levels in the quantum well. Subsequent evolution of this
model takes into account inelastic scattering in the quantum well by
means of an inelastic contribution to the effective width of the
energy range of well states, which broadens the resonance and
lowers the peak of the tunneling probability as a function of
energy.\cite{stonelee85,jonsgrin87}

The other well known model is that of ``sequential tunneling,''
proposed by Luryi,\cite{luryi85} which consists in considering
tunneling through the double barrier as a two-step process:
first, the transition from one electrode into the well, then
the transition from the well to the other electrode.
The two sequential transitions are assumed to occur incoherently,
so that scattering in the well region is accounted for implicitly;
moreover, the amount of inelastic scattering in the well
influences the width of the density of states in the well as
a function of energy.

Zohta \cite{zohta93} has pointed out some ambiguity in the
definition of sequential tunneling. In fact, in some papers
sequential tunneling is intended as a transport mechanism, i.e.,
what happens when the electron
traverses one barrier coherently, is scattered inelastically in
the well and loses phase memory, then escapes through the other
barrier;\cite{buettiker88} in other
papers sequential tunneling is considered just as an alternative
way to describe the resonant tunneling phenomenon.\cite{zohta93,weilvint87}

In order to clarify these ambiguities, in this paper we make a clear
distinction between approaches and transport regimes. Here we
discuss two different approaches to the study of transport in double
barrier structures: the ``resonant tunneling'' one, and the
``sequential tunneling'' one.
On the other hand, we have a whole range of transport regimes,
depending of the rate of collision processes in the well, i.e., on the
value of the effective mean free path $l$: one limit is the
completely coherent transport ($l \rightarrow \infty$),
when no collisions occur in the
well, and the other limit is completely incoherent transport
($l \rightarrow 0$), when
the high collision rate cancels out any size effect in the well.

We wish to point out that, in this paper, the word ``sequential''
is used to denote both a transport
regime and an approach, in order to follow the usual terminology adopted
in the literature. \cite{luryi85,zohta93,buettiker88,weilvint87}
This fact should not be misleading to the reader: while the
sequential ``regime,'' discussed in Sec.IVB,
is a well defined transport regime in which almost all electrons
lose phase coherence in the well, \cite{luryi85,buettiker88}
the sequential tunneling ``approach'' is just the method of
considering the double barrier structure as consisting of
three isolated regions weakly coupled through the barriers (which
have to be opaque so that Bardeen tunneling Hamiltonian\cite{bardeen61}
can be used to obtain transition matrix elements).

As such, the sequential
approach does not require loss of phase coherence in the well to be
applicable; conversely, it can be used to describe also coherent
transport. Of course, the amount of scattering in the well is important
if we use the sequential tunneling approach: in fact, it affects the
density of states in the well \cite{zohta93,iannaora}
which, in turn, affects the number of
transitions per unit time between the well and any of the electrodes
through the Fermi golden rule.

In section III we show that the approaches mentioned above,
properly extended, are completely equivalent and either
can be used to describe all
possible transport regimes. This result has been obtained
by Veil and Winter in the case of
completely coherent transport,\cite{weilvint87}
and by Zohta in the case of simmetric barriers and inelastic
scattering in the well.\cite{zohta93}
We extend this equivalence to the
general case of arbitrary barriers, and arbitrary amount
of scattering in the quantum well, so that the whole range
of transport regimes is addressable.
The derivation of this equivalence, as we shall show, is based on
a recently derived formula for the density of states in the
well region of a double barrier structure.\cite{iannaora}

In section IV we discuss three relevant transport regimes:
the ``coherent,'' ``sequential,'' and ``completely incoherent''
regimes, for all of which, as said above, either approaches
can be used.
However, for the coherent regime, where the prevalent contribution to
the total current comes from electrons which conserve phase coherence
and energy, the resonant tunneling approach is more straightforward.
Regarding the sequential regime,
the sequential tunneling approach is the best suited: in this regime
practically all electrons lose phase memory and thermalize in the well,
but size effects strongly affect the density of states in the well.
Finally, also in the completely incoherent regime, where
size effects are cancelled out so that
the density of states in the well approaches that
in the bulk, the sequential tunneling approach is the most direct one.

\section{Transport model}

Transport properties of ultrasmall structures strongly depend on
both elastic scattering (due to impurities, crystal defects, and
interface roughness) and inelastic scattering (due to
phonons and electron-electron interactions).
Elastic collisions conserve energy and phase coherence,
while inelastic collisions do not.

However, if there is a sufficiently large number of impurities and
defects randomly distributed, also elastic scattering has
phase randomizing effects, due to averaging over many paths
corresponding to different actions.\cite{zohta89,hersambe88,laugbark91}
Therefore, we can account for dephasing effects of both elastic
and inelastic collisions by means of a single characteristic length
$l$, the effective mean free path, which is a phenomenological parameter
and, as we shall see, plays the role of a relaxation length.

We assume that an electron traversing
a length $dx$ of the one-dimensional device structure has
a probability $dx/l$ of experiencing
a collision, and that electrons emerge from collisions with a
quasi-thermal equilibrium energy distribution and a completely random phase,
so that there is no quantum interference between these electrons
and the ones which have not
lost phase coherence.

As can be seen, energy relaxation and phase randomization processes are
supposed to occur at the same time, and to be triggered by any kind of
collision,
regardless it is elastic or inelastic.
This is not rigorously true: in fact, electrons do not lose energy in
elastic collisions, and inelastic collisions are presumably not so
effective in relaxing energy as they are in randomizing phase. Anyway we shall
use, for simplicity, a unique lenght scale for both processes.
A more sophysticated model should distinguish between different
scattering sources and incohence phenomena. For instance,
two different energy-dependent characteristic lengths for
phase randomization and energy relaxation could be adopted.

Our model is close to the one proposed by
B\"uttiker;\cite{buettiker88,buettiker2_86}
however B\"uttiker's model is applicable when the differences
between electrode chemical potentials are small and/or when
energy relaxation is not accounted for. Moreover, in our model
scattering phenomena can be spread---in principle---over the
whole region of interest, and are not concentrated in a single
inelastic scatterer coupling the well to an extra reservoir.

The idea of using a phenomenological mean free path to account for
the loss of phase coherence in the well is not new.
\cite{zohta93,zohta89,hersambe88,hu88,zohta90,knabchen92}
As an improvement to previous similar models,
we have associated to the effective mean free path also
energy relaxing processes: in the following, we shall show that
this step is important to take into
account dissipation and to obtain the main result of this paper,
i.e., the demonstration of the equivalence of the sequential
and resonant tunneling approaches.

Other characteristic lengths usually considered in the study of tranport
in ultrasmall structures are not relevant to our discussion: all
currents in the following sections are obtained after integration over
all energies, so that thermal averaging due the to spreading of the
Fermi-Dirac
distribution at non-zero temperatures is already taken into account,
making unnecessary the evaluation of
the thermal diffusion length $l_T$. \cite{vanhouten89} Moreover,
double barrier structures are more extended in the transverse plane than
in the longitudinal direction, therefore localization effects are
negligible (and the localization length is not relevant).

As we shall show,
using a single effective mean free path for taking into account the
effects of both the elastic and inelastic mean free paths,
transport regimes in the considered structures
depend only on the relations between $l$,
the well width $w$, and the barrier transmission probabilities $T_1$
and $T_2$.

\subsection{Double barrier structure}

We refer to the system shown in Fig.\ 1. The potential $V(x)$
defines two barriers and the well region. Given the electron
longitudinal energy $E$, we can use the transfer matrix
technique\cite{erdohern83} and the multistep potential
approximation\cite{andoitoh87} to calculate the transmission and
reflection probabilities for the whole structure and for each single
barrier.

Scattering in the well can be easily accounted for by using in the
transfer matrices the complex wave vector $k_i(x) = k(x) + i/2l$, where
$k(x) = [2m(E-V(x))]^{1/2}/\hbar$, and $m$, $\hbar$
are the electron effective mass in the material of the well and
the reduced Planck's constant, respectively.

Transport in the single barriers is assumed to be completely
coherent; therefore,
if $T_i$ and $R_i$ are the tunneling and reflection probabilities for barrier
$i$ ($i = 1,2$), we have $T_i + R_i = 1$. On the other hand, collisions
in the well make the continuity equation for the probability density
current of a given state no longer applicable, and we have $T_{db}^{l} +
R_{db}^l < 1$, where $T_{db}^l$ and $R_{db}^l$ are the transmission
and reflection probabilities of the whole double barrier structure
for an electron coming from the left electrode (their expression is derived in
Ref. \onlinecite{iannaora}). The same relation applies to $T_{db}^r$ and
$R_{db}^r$, where the superscript $r$ stays for the right electrode.

In the absence of magnetic field, from time reversal symmetry we
have $T_{db} \equiv T_{db}^l = T_{db}^r$, while in general
$R_{db}^l \neq R_{db}^r$.\cite{timereversal}
By means of a general relation between the density of states and dwell
times in mesoscopic systems \cite{ianna95}, in Ref. \onlinecite{iannaora}
we obtained that, on the assumption
of smooth potential in the well region, the density of
states $\rho_w(E)$ in the effective well region (i.e., including
states in the well and tail states penetrating
the well sides of both barriers)
can be written as
\begin{equation}
\rho_w(E) = \frac{1}{ \pi \hbar}
		\tau_{rt}^i
	\left[ \frac{1 - R_{db}^l}{T_1} + \frac{T_{db}}{T_2} \right]
,\label{doswbella}
\end{equation}
where both spin components have been considered and
$\tau_{rt}^i$ is the round trip time in the well at the resonant enengy.
In the following sections we shall show the importance of this formula in
unifying the resonant tunneling and the sequential tunneling approaches.

\section{Approaches to transport in double barrier structures}

\subsection{Resonant tunneling approach}

The introduction of the effective mean free path $l$ has the effect that the
current probability density for a given state is not conserved. Electrons
which experience a collision
seem to be ``absorbed'' in the well; in fact, they actually emerge
with a quasi-thermal equilibrium distribution, for which we need
to introduce the density of states in the well and a quasi-Fermi energy
level.

Let $dJ_1$ be the contribution to the current through the first barrier
due to electrons with longitudinal
energies between $E$ and $E + dE$, referred to the
conduction band bottom of the left electrode.
We can write $dJ_1$ as the sum of three terms, i.e.
\begin{equation}
dJ_1 = dJ_1^l + dJ_1^w + dJ_1^r
,\label{djdec}
\end{equation}
where $dJ_1^l$, $dJ_1^w$, and $dJ_1^r$ are the current contributions
due to electrons which have sufferred their latest collision
(and have emerged with equilibrium energy distribution from)
the left region, the well, and the right region, respectively.
Let ${\bf k_T}$ be the transverse wave vector, and $\rho_T({\bf k_T})$
the density of transversal states. Moreover, let us indicate
with $\rho_s(E)$, $f_s(E,{\bf k_T})$, $\nu_s(E)$, ($s=l,w,r$),
the one-dimensional density of states for
longitudinal energies (including both spin contributions),
the occupation factor, and
the attempt frequency (i.e., the average number of bounces on
each barrier per second), respectively: the subscripts $l$, $w$, and $r$
refer the left electrode, the well, and the right electrode,
respectively. We also introduce the
quasi-equilibrium occupation factor in the well $f_{w0}(E,{\bf k_T})$
corresponding to the quasi-Fermi energy $E_{fw0}$.
We can define the integral of the occupation factor over
transversal wave vectors $F_s(E)$ ($s=l,w,r,w0$), as
\begin{equation}
F_s(E) = \int f_s(E,{\bf k_T}) \rho_T({\bf k_T}) d{\bf k_T}
,\label{fdis}
.\end{equation}

Therefore we can write
\begin{eqnarray}
dJ_1^l & = & q (1-R_{db}^{l}) \rho_l F_l \nu_l dE
,\label{dj1l} \\
dJ_1^w & = & -q T_1 \rho_w F_{w0} \nu_w dE
,\label{dj1w} \\
dJ_1^r & = & -q T_{db} \rho_r F_r \nu_r dE
,\label{dj1r}
\end{eqnarray}
where we avoid to write explicitly the dependence of all the terms
upon the longitudinal energy $E$.
We also assume that the contacts are ``ideal,'' in the sense
that they absorb without reflection all electrons leaving the device,
and inject electrons according to the thermal equilibrium distribution.
This modeling is implicitly assumed when transport in quantum devices
is described as a scattering event, \cite{buettiker86} and,
as long as we deal with stationary regimes, is applicable without any
restriction.

The longitudinal density of states in the left region can be calculated
with periodic boundary conditions as \cite{longdos}
\begin{equation}
\rho_l(E) = \frac{L_l}{\pi \hbar v_l(E)}
\label{rhol}
,\end{equation}
where $L_l$ is the length of the left electrode and $v_l(E)$ is the
longitudinal velocity corresponding to $E$. The attempt frequency
(corresponding to periodic boundary conditions) is simply
$ \nu_l(E) = v_l(E)/L_l$; so we have $\rho_l(E) \nu_l(E) = 1/\pi \hbar$.
The same considerations apply to the
right region, therefore we have $\rho_r(E) \nu_r(E) = 1/\pi \hbar$.
The attempt frequency in the well is just the inverse of the
time required to complete a round trip of the well, i.e.,
$\nu_w(E) = 1 /\tau_{rt}^i(E)$.

We can write the expression for the current through the
second barrier in a similar way.
The total current through barrier $i$ ($i=1,2$) is $J_i = \int dJ_i$.
In stationary conditions we must have
\begin{equation}
J_1 = J_2
.\label{jcont}
\end{equation}
By imposing this equality we can obtain the quasi-Fermi level
$E_{fw0}$ in the well.

\subsection{Sequential tunneling approach}

According to this approach,
electron tunneling through the double barrier is considered as a
two-step process, following Weil and Vinter's formalism.
\cite{zohta93,weilvint87}
If the tunneling probabilities of the two barriers are very small, \cite{}
i.e., $T_1, T_2 \ll 1$,\cite{sakurai85}
we can write $d(J_{1})'$ and $d(J_2)'$,
the contributions to the current through
barriers 1 and 2 due to
electrons with energies in the interval $(E, E+dE)$, as
\begin{equation}
d(J_{1})' = \frac{\pi q}{\hbar} | M_1 |^2 \rho_l \rho_w (F_l - F_w) dE,
\label{dij1}
\end{equation}
and
\begin{equation}
d(J_{2})' = \frac{\pi q}{\hbar} | M_2 |^2 \rho_w \rho_r (F_w - F_r) dE
,\label{dij2}
\end{equation}
where $M_1$ ($M_2$) is the matrix element for the transition from
a state in the left (right) region
to a state in the well.\cite{bardeen61}
We have (detailed derivation is shown in appendix A)
\begin{eqnarray}
M_1 & = & \hbar^2 T_1 \nu_l \nu_w,\\
M_2 & = & \hbar^2 T_2 \nu_w \nu_r
.\end{eqnarray}

In the original model, \cite{luryi85}
the two transitions are assumed to occur incoherently, therefore energy
relaxation and phase randomization in the well are implicitly accounted
for. We can extend this model in order to include the effects of
ballistic electrons
and of coherent transport by simply considering the occupation factor
$f_w$ in the well as a superposition of three partial occupancy
probabilities,
a quasi-equilibrium one, $f_{w0}$, and the occupancy probabilities
$f_w^r$ and $f_w^l$ for electrons coming from the left and right electrodes,
which have not been randomized, i.e.,
\begin{equation}
f_w = f_{w0} + f_w^l + f_w^r
.\label{fwell}
\end{equation}
If we substitute (\ref{fwell}) in (\ref{fdis}) and then in
(\ref{dij1}), we can write
the current contribution $d(J_1)'$ of electrons with
longitudinal energies between $E$ and $E+dE$ as (\ref{djdec}),
provided that we choose
\begin{eqnarray}
d(J_1^l)' & = & q \frac{\pi}{\hbar} | M_1 |^2 \rho_l \rho_w ( F_l - F_w^l) dE
,\label{dj1lseq} \\
d(J_1^w)' & = & -q \frac{\pi}{\hbar} | M_1|^2 \rho_l \rho_w F_{w0} dE
,\label{dj1wseq} \\
d(J_1^r)' & = & -q \frac{\pi}{\hbar} | M_1|^2 \rho_l \rho_w F_w^r dE
.\label{dj1rseq}
\end{eqnarray}
In $d(J_1^l)'$, $d(J_1^w)'$, and $d(J_1^r)'$ we have taken into account
the contribution of
electrons emerging with equilibrium distribution from the
left, well, and right region, respectively.

\subsection{Equivalence between the above approaches}

In order to verify the equivalence between the resonant tunneling
and sequential tunneling approaches we have to make explicit
the conditions that guarantee that the
values of $dJ_1^s$, $(s = l,w,r)$ of (\ref{dj1l}), (\ref{dj1w}), and
(\ref{dj1r}) be equal to $d(J_1^s)'$ of (\ref{dj1lseq}), (\ref{dj1wseq})
and (\ref{dj1rseq}), respectively, and that the same relations
apply to the corresponding terms for $dJ_2$ and $d(J_2)'$.
Straightforward calculations yield the required conditions:
\begin{equation}
n_w^l \equiv \rho_w F_w^l = \frac{1}{\pi \hbar \nu_w} \frac{T_{db}}{T_2} F_l
,\label{nwl}
\end{equation}
\begin{equation}
n_w^r \equiv \rho_w F_w^r = \frac{1}{\pi \hbar \nu_w} \frac{T_{db}}{T_1} F_r
,\label{nwr}
\end{equation}
and
\begin{equation}
\rho_w = \frac{1}{\pi \hbar \nu_w}
	\left[ \frac{1-R_{db}^l}{T_1} + \frac{T_{db}}{T_2} \right]
	\approx
	\frac{1}{\pi \hbar \nu_w}
        \left[ \frac{1-R_{db}^r}{T_2} + \frac{T_{db}}{T_1} \right]
.\label{dos}
\end{equation}
As long as (\ref{nwl}-\ref{dos}) hold true, the two
models can be thought as completely equivalent.
It is straightforward to see that conditions
(\ref{nwl}) and (\ref{nwr}) are satisfied:
in fact, given $\rho_l \nu_l = 1/\pi \hbar$,
we can re-write (\ref{nwl}) as
\begin{equation}
n_w^l \nu_w T_2 = \rho_l F_l \nu_l T_{db}
,\label{bibo}
\end{equation}
that can be read as follows: $\rho_l F_l$ is the number of
electrons in the left region, $\nu_l$ the bounces on the first barrier per
second, $T_{db}$ is the coherent tunneling probability,
therefore the right term is the number of electrons per second coehently
traversing
the double barrier. The left term has the
same meaning, given that $n_w^l$ is the number of electrons
in the well which have come from the left region and have not
lost phase coherence.

The condition given by (\ref{dos}) is exactly equal to (\ref{doswbella})
which has been obtained in Ref. \onlinecite{iannaora},
therefore the equivalence of the two approaches has been demonstrated.

\section{Transport regimes through the double barrier}

Now, according to (\ref{djdec}-\ref{rhol}) and (\ref{dos}), we have
\begin{eqnarray}
J_1 & = & \frac{q}{\pi \hbar}  \int
\left[ (1 - R_{db}^l)F_l - \left( 1 - R_{db}^l + \frac{T_1T_{db}}{T_2}
        \right) F_{w0} - T_{db}F_r \right] dE
\label{j1lunga}\\
J_2 & = & \frac{q}{\pi \hbar} \int
\left[ T_{db} F_l + \left( 1 - R_{db}^r + \frac{T_2 T_{db}}{T_1}
        \right) F_{w0} - (1 - R_{db}^r) F_r \right] dE
.\label{j2lunga}
\end{eqnarray}
With these results, following the scheme by
B\"uttiker \cite{buettiker88},
we can discuss three relevant transport
regimes in double barrier structures,
depending on the amount of inelastic scattering in the
well: the coherent transport regime,
the sequential regime, and the completely incoherent regime.

\subsection{Coherent transport regime}

In this regime the prevalent contribution to the total current
comes from ``ballistic'' electrons, which do not lose phase
coherence and energy in the well. This happens if
the probability that a particle suffer
from collisions in the well is close to zero, i.e.,
\begin{equation}
T_{db} \approx 1 - R_{db}^l \approx 1 - R_{db}^r
.\label{cohecond}
\end{equation}
This condition is satisfied if the
effective mean free path is long enough that almost any
particle escapes from the barrier before undergoing scattering
events, in other words $w/l \ll \min\{T_1,T_2\}$.
Substitution of (\ref{cohecond}) in (\ref{j1lunga}-\ref{j2lunga})
easily yields $F_{w0} = 0$, as it has to be, and
\begin{equation}
J_1 = J_2 = \frac{q}{\pi \hbar} \int_0^{\infty}
                        T_{db} \left( F_l - F_r \right) dE
.\label{cohecurr}
\end{equation}
Therefore, according to this result and to (\ref{dj1l}-\ref{dj1r}),
the resonant tunneling approach suits very well this transport
regime, by simply taking
$dJ_1^w = dJ_2^w = 0$, $dJ_1^l = dJ_2^l$, $dJ_1^r = dJ_2^r$.

\subsection{Sequential transport regime}

We call sequential regime the situation in which practically all
electrons suffer from collisions before escaping from the
well, i.e.,
\begin{equation}
T_{db} \ll 1 - R_{db}^l, \hspace{0.5cm} T_{db} \ll 1 - R_{db}^r
,\label{sequecond}
\end{equation}
and, however, they complete at least a few round trips of the well
before escaping, so that the density of states $\rho_w$
is affected by the confinement. As can be seen from (5-7) of
Ref. \onlinecite{iannaora}, it means that $T_1,T_2 \ll w/l <1$,
and implies that both $T_{db} T_2/T_1$ and $T_{db} T_1/T_2$ are
much smaller than either $1-R_{db}^l$ or $1-R_{db}^r$.
Taking into account (\ref{sequecond})
in (\ref{j1lunga}-\ref{j2lunga}) we obtain
\begin{eqnarray}
J_1 & = & \frac{q}{\pi \hbar} \int (1 - R_{db}^l)(F_l -F_{w0}) dE
\label{sequej1},\\
J_2 & = & \frac{q}{\pi \hbar} \int (1 - R_{db}^r)(F_{w0} -F_r) dE
\label{sequej2}
.\end{eqnarray}
Moreover, the density of states in the well given by (\ref{dos}) simply
becomes $\rho_w \approx (\pi \hbar \nu_w T_1)^{-1} (1-R_{db}^l)$, and
from (\ref{nwl}) and (\ref{nwr}) we readily verify that
ballistic electron concentrations $n_w^l$ and $n_w^r$
in the well are almost zero. As
can be seen from (\ref{sequej1}) and (\ref{sequej2}),
the double barrier traversal becomes essentially a two-step process,
and any effect of quantum interference is implicitly included
in $R_{db}^r$ and $R_{db}^l$.
The sequential tunneling approach suits directly this
regime, by simply substituting $F_w$ with $F_{w0}$ in (\ref{dij1})
and (\ref{dij2}).

\subsection{Completely incoherent transport regime}

For higher rates of incoherent processes, i.e., $w/l \gg 1$,
we have that
\begin{equation}
1 - R_{db}^l \approx T_1, \hspace{0.5cm} 1 - R_{db}^l \approx T_2,
\hspace{0.5cm} T_{db} \approx T_1 T_2
,\label{incocond}
\end{equation}
as can be seen from (6) of Ref. \onlinecite{iannaora}. Therefore
(\ref{sequej1}) and (\ref{sequej2}) become
\begin{equation}
J_1 = \frac{q}{\pi \hbar} \int T_1 (F_l -F_{w0}) dE
,\label{incoj1}
\end{equation}
\begin{equation}
J_2 = \frac{q}{\pi \hbar} \int T_2 (F_{w0} - F_r) dE
;\label{incoj2}
\end{equation}
Correlation between currents through barriers 1 and 2 is only
due to the current conservation in the well, i.e., to the
position of the quasi-Fermi level in the well.
The density of states in the well, as can be seen from (\ref{dos}),
becomes $\rho_w \approx (\pi \hbar \nu_w )^{-1}$, i.e.,
any size effect disappears, and $\rho_w$ is equal to the
density of states in the bulk of the material given by (\ref{rhol}).
The sequential tunneling approach suits easily this regime,
by simply putting $F_w = F_{w0}$ and $\rho_w \nu_w = (\pi \hbar)^{-1}$
in (\ref{dij1}) and (\ref{dij2}).

\section{Summary}

In this paper we have shown that the sequential tunneling and
resonant tunneling approaches, properly extended, are completely
equivalent for any rate of collision processes in the well,
and can be used to describe all the range of
transport regimes in double barrier structures. Let us point
out again the extensions required.

In the resonant tunneling approach we account for
incoherent transport by introducing a current due
to electrons scattered in the well and emerging with equilibrium
distribution (namely $dJ_1^w$ and $dJ_2^w$). Determining
this current requires knowledge of the density of states and
of the occupancy probability of the states in the well.

In the sequential tunneling description we account for
electrons which have not been scattered in the well by
means of the  ``ballistic'' distributions $n_w^l$ and $n_w^r$.
Also in this description the density of states and the
equilibrium distribution function are fundamental parameters.

The difference between these approaches is in the point
of view which is given preference: the former
assumes that electrons (except the ones that are inelastically scattered)
traverse the structure coherently, and that the latter assumes
that electrons (except ballistic ones)
thermalize in the well and obey to an equilibrium energy distribution.
Therefore, it is apparent that the resonant tunneling approach
is better suited for describing coherent transport, and
the sequential tunneling approach fits more directly to incoherent
transport, when practically every electron undergoes inelastic
scattering in the well.

The equivalence of these two models, for the whole range of
transport regimes, has been shown on the
basis of a recently derived formula for the density of
states in a quantum well as a function of the round trip time
and of transmission and reflection probabilities for
the whole structure and for each barrier. Energy relaxation
and phase-breaking phenomena are accounted for, in our
simple model, by means of a single parameter, the effective mean free
path $l$, which plays the role of a relaxation length.

We wish to point out that while
in the paper by Zohta \cite{zohta93} this equivalence was verified only
for symmetric barriers, not accounting for energy relaxation
in the well, and as long as the Breit-Wigner
formalism was applicable, here the equivalence between resonant
tunneling and sequential tunneling approaches
is demonstrated
for any double barrier structure and for any degree of
inelastic scattering, considered as a randomizing agent for both
phase and energy.

\section{Acknowledgments}

The present work has been supported by the Ministry for the University
and Scientific and Technological Research of Italy, by the Italian
National Research
Council (CNR). The authors wish to thank M. Macucci for helpful
suggestions and M. B\"uttiker for useful critical comments to the manuscript.

\appendix

\section*{Matrix element for the transition through a potential barrier}

Let us consider the one-dimensional potential barrier sketched in
Fig. 2, separating regions 1 and 2. The potential energy around the
turning points $a$ and $b$ has been modified in order to have two flat
steps, so that the wave functions on both sides of the potential
barrier can be written as a superposition of plane waves. In fact
we use a perturbed $V'(x)$ defined as
\begin{equation}
V'(x) = \left\{
\begin{array}{ccc}
V(a - \epsilon), & \hspace{0.5cm} & a-\epsilon < x < a + \epsilon \\
V(b + \epsilon), & & b-\epsilon < x < b + \epsilon \\
V(x) & & \mbox{otherwise}
\end{array}
\right.
.\end{equation}
However, the step width $2\epsilon$ can be arbitrarily small, therefore we do
not loose generality.

Around $a$ and $b$, the wave function $\psi_1$ of an electron of
energy $E$ coming from region 1 is
\begin{equation}
\psi_1 = \left\{
\begin{array}{lcl}
A_1 [ e^{ik_1(x-a)} + r^l e^{-ik_1(x-a)} ],&& a-\epsilon <x<a+\epsilon
\\
A_1 t^l e^{ik_2(x-b)},	&& b - \epsilon<x<b+\epsilon
\end{array}
\right.
,\label{psi1}
\end{equation}
where $k_1= [2m(E-V'(a)]^{1/2}/\hbar$, $k_2 = [2m(E-V'(b)]^{1/2}/\hbar$,
and $r^l$, $t^l$ are the
reflection and transmission coefficients of the barrier for an electron
coming from the left. If $\psi_2$ is the wave function of an electron
at the same energy in region 2 we can write
\begin{equation}
\psi_2 = \left\{
\begin{array}{lcl}
A_2 t^r e^{-ik_1(x-a)}, && a-\epsilon <x<a+\epsilon \\
A_2 [e^{-ik_2(x-b)} + r^r e^{i k_2(x-b)} ], && b - \epsilon<x<b+\epsilon
\end{array}
\right.
,\label{psi2}
\end{equation}
where $r^r$ and $t^r$ are the reflection and transmission coefficients
for an electron coming from the right. From Bardeen \cite{bardeen61}
we can calculate the matrix element from the transition between
regions 1 and 2 at energy $E$, as
\begin{equation}
|M|^2 = \frac{\hbar^4}{4 m^2}
 \left. \left| \psi_1^* \nabla \psi_2 - \psi_2 \nabla \psi_1^*
\right|^2 \right|_{x=a}
 =  \hbar^2 T J_{1inc} J_{2inc}
\end{equation}
where $T = |t^r|^2 k_2/k_1$ is the transmission probability of the
barrier;
$J_{1inc}$ ($J_{2inc}$) is the probability current incident
on the barrier associated to the state $\psi_1$ ($\psi_2$); for
instance, from (\ref{psi1}) we find that $J_{1inc} = |A_1|^2 \hbar k_1/m$;
however, $J_{1inc}$ is simply obtained as
the integral of the probability density
in region 1 times the attempt frequency on the barrier; i.e.,
if $\psi_1$ is normalized to unity, $J_{1inc} = \nu_1$. If we
apply the same result to $J_{2inc}$ we can eventually write
\begin{equation}
|M|^2 = \hbar^2 T \nu_1 \nu_2
.\end{equation}

\begin{figure}
\vspace{2cm}

\epsfxsize=\figwidth
\epsffile{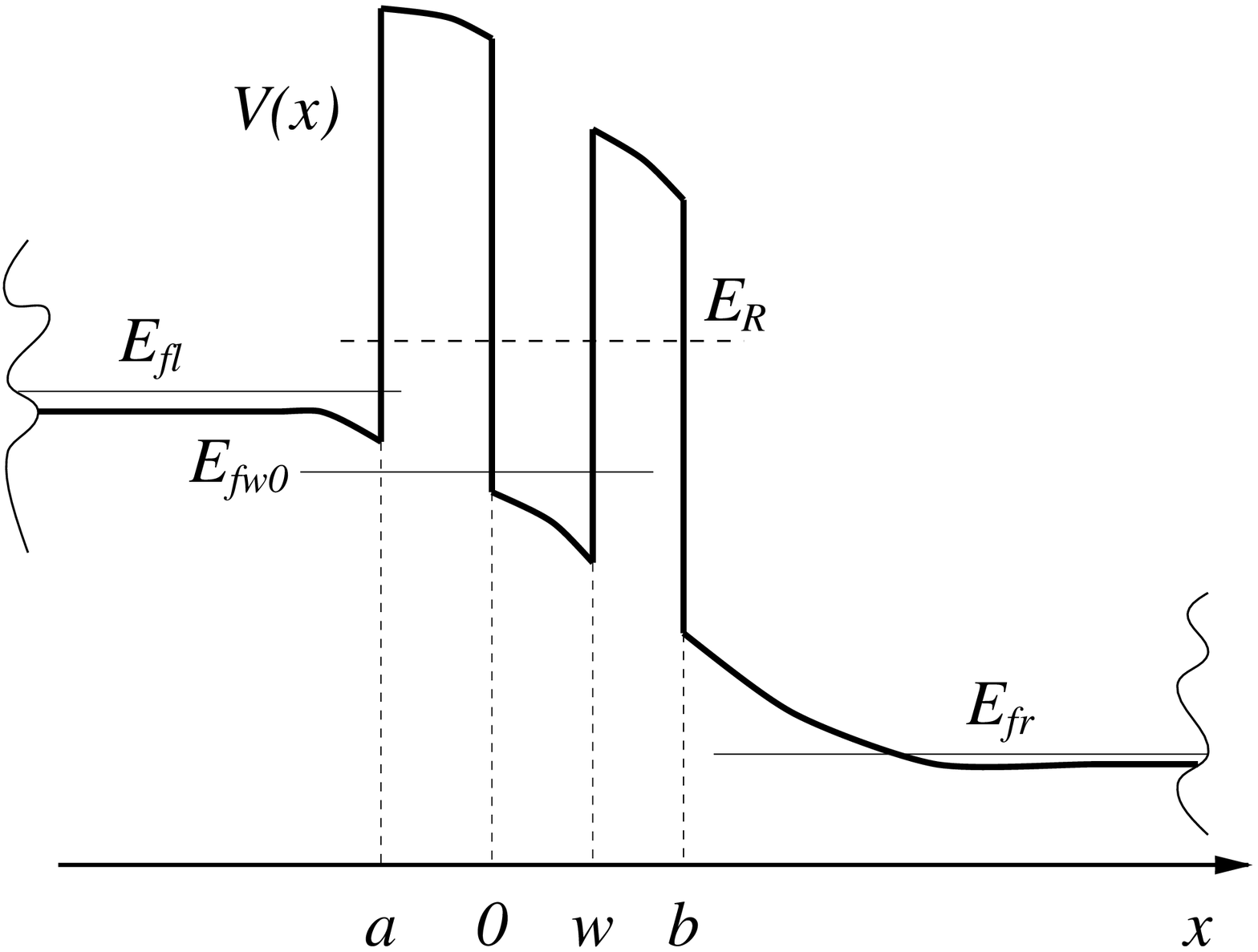}

\vspace{1cm}

\caption{The one-dimensional potential energy profile $V(x)$
defines the first barrier $(a,0)$, the well region $(0,w)$, and
the second barrier $(w,b)$. $E_{fl}$, $E_{fw0}$, $E_{fr}$ are
the quasi-Fermi levels in the left electrode, well, and right
electrode, respectively.}
\end{figure}

\begin{figure}
\epsfxsize=\figwidth
\epsffile{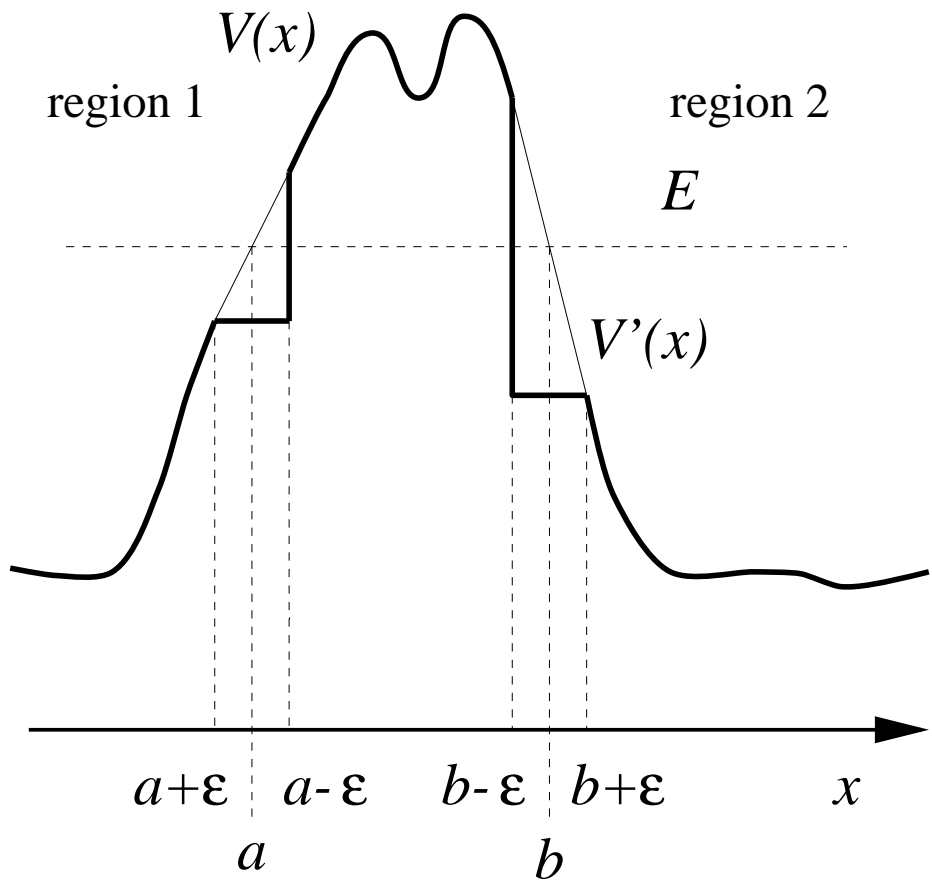}

\vspace{1cm}

\caption{The one-dimensional potential energy $V(x)$ (fine line)
separates regions 1 and 2. The potential energy profile $V'(x)$
(thick line) is equal to $V(x)$ except near the turning points $a$ and $b$,
where is has been modified in order to have flat steps, so that the
wave functions on both sides of the potential barrier can be written
as a superposition of plane waves: $V'(x) = V(a-\epsilon)$ for
$a-\epsilon<x<a+\epsilon$, and $V'(x) = V(b+\epsilon)$ for
$b-\epsilon<x<b+\epsilon$.}
\end{figure}
\end{document}